\documentclass[conference]{ieeeconf}  

\IEEEoverridecommandlockouts
\usepackage{amsmath,amssymb,amsfonts} 

\newcommand{\fullversion}{\color{black}}
\usepackage{algorithm}
\usepackage{algpseudocode}
\usepackage{graphicx}
\usepackage{color}
\usepackage{booktabs}
\usepackage{cite}

\usepackage[hyphens]{url}
\usepackage{breakurl}
\usepackage{makecell}
\usepackage{multicol}
\usepackage{float}
\usepackage{subfig}

\begin{document}

\title{\LARGE \bf Model Predictive Building Climate Control for \\Mitigating Heat Pump Noise Pollution (Extended Version)}

\author{Yun Li$^{1}$, Jicheng Shi$^{2}$, Colin N. Jones$^{2}$, Neil Yorke-Smith$^{3}$, and Tamas Keviczky$^{1}$
\thanks{The work was supported by the Brains4Buildings project under the Dutch grant programme for Mission-Driven Research, Development and Innovation (MOOI), and the Swiss National Science Foundation (SNSF) under the NCCR Automation project, grant agreement 51NF40\_180545.}
\thanks{$^{1}$Yun Li and Tamas Keviczky are with Delft Center for Systems and Control, Delft University of Technology,  Delft, the Netherlands. {\tt\small y.li-39@tudelft.nl, T.Keviczky@tudelft.nl}}
\thanks{$^{2}$Jicheng Shi and Colin N. Jones are with Automatic Control Laboratory, EPFL Lausanne, Switzerland. {\tt\small jicheng.shi@epfl.ch, colin.jones@epfl.ch}}
\thanks{$^{3}$Neil Yorke-Smith is with STAR Lab, Delft University of Technology, Delft, the Netherlands. {\tt\small n.yorke-smith@tudelft.nl}}
}

\maketitle

\begin{abstract}
Noise pollution from heat pumps (HPs) has been an emerging concern to their broader adoption, especially in densely populated areas. This paper explores a model predictive control (MPC) approach for building climate control, aimed at minimizing the noise nuisance generated by HPs. By exploiting a piecewise linear approximation of HP noise patterns and assuming linear building thermal dynamics, the proposed design can be generalized to handle various HP acoustic patterns with mixed-integer linear programming (MILP). Additionally, two computationally efficient options for defining the noise cost function in the proposed MPC design are discussed. Numerical experiments on a high-fidelity building simulator are performed to demonstrate the viability and effectiveness of the proposed design. Simulation results show that the proposed approach can effectively reduce the noise pollution caused by HPs with negligible energy cost increase.
\end{abstract}

\section{Introduction}
As an energy-efficient heating/cooling device, heat pumps (HPs) have gained widespread adoption across Europe, driven by the goal of reducing fossil fuel usage and carbon emissions. Currently, approximately 24 million HPs are installed in European buildings, and this number is expected to reach 60 million by 2030 \cite{ehpa2024}. This growing adoption of HPs to move away from fossil fuels could reduce Europe's gas demand for heating by at least 21 billion cubic meters in 2030, and potentially cut CO2 emissions by 46\% \cite{ec23,guardian24}. However, despite the benefits of flexible and efficient renewable heating and carbon reduction, a new concern about HPs has been raised: noise. 

Noise can induce stress and impact both psychological and physiological well-being. Noise generated by HPs, particularly air source heat pumps (ASHPs) commonly installed in residential areas, has emerged as a primary concern hindering their broader acceptance in these settings \cite{langerova2025air,M20}. Consequently, HP installation and operation must account for acoustic impacts on the surrounding environment, especially in residential zones where noise levels are subject to legislative noise directives. For example, in the UK, the noise pressure level must be below 42 dB at a distance of one meter from a neighbour's door or window \cite{M20,guardian24}. Similar regulations also apply in other countries as outlined in \cite{ehpa20}. Addressing HP noise pollution is thus essential for maintaining acoustic health and fostering the acceptance of HPs, which can further support carbon emissions reduction. Recent initiatives, such as IEA HPT Annex 51 and Annex 63 underscore the growing attention on HP noise concerns \cite{Annex51,annex63}.

Various solutions for reducing HP noise have been explored, including adding sound-absorbing materials or insulation enclosures, using flexible mountings to dampen vibrations, and implementing active noise cancellation techniques \cite{wag20,langerova2025air,thielecke2023active}. While these measures can reduce HP noise, they often require intrusive modifications, making it costly or even impractical to retrofit existing HPs for improved acoustic performance. 

With advancements in smart metering, computing technology, and building management systems, an alternative approach to reducing HP noise without invasive modifications is to design optimal HP control strategies. The primary noise sources of air source heat pumps are the compressors and, especially, the fans in the outside units of HPs, which significantly contribute to ambient noise \cite{M19,langerova2025air}. Modern inverter HPs allow for modulation of compressor and fan speeds, enabling noise level adjustments based on thermal output requirements. Thus, it is possible to adjust the HP power inputs to mitigate noise while maintaining a comfortable indoor thermal climate. Model predictive control (MPC) has shown promise as an advanced control strategy for building climate control, owing to its flexibility in handling system constraints, economic considerations, and predicted weather conditions \cite{oldewurtel2012use,drgovna2020all}.


Motivated by the above discussion, this paper investigates an MPC design to mitigate HP noise within the context of building climate control. To the best of our knowledge, this is the first study to address the noise nuisance from HP operation by designing optimal building climate control schemes. The main contributions of this paper are summarized as:
\begin{itemize}
    \item The reduction of HP noise pollution is investigated for the first time in optimal building climate control. A general MPC formulation that considers both HP noise pollution reduction and energy cost savings is proposed.
    \item Leveraging piecewise linear approximation, the proposed design is adaptable to various HP noise patterns through a mixed-integer linear programming (MILP) formulation. Two options for the noise cost function in the MPC design are discussed.
    \item Numerical experiments using a high-fidelity building simulator are performed to demonstrate the viability and effectiveness of the proposed approach.
\end{itemize}

The remaining parts of this paper are organized as follows. Section \ref{sec:problem_setting} presents the problem setting about building thermal dynamics, HP noise patterns, and our design objective. Section \ref{sec:mpc_design} delves into the details of the proposed MPC design. The viability and effectiveness of the proposed approach are numerically tested in Section \ref{sec:simu_results}. Finally, Section \ref{sec:conclusion} concludes this paper.

\section{Problem Setting}\label{sec:problem_setting}
\subsection{Building Thermal Dynamics}
Without loss of generality, the indoor thermal dynamics are assumed to be approximately modeled as the following linear system
\begin{equation}\label{eq:thermal_dyn}
    {y}_{t+1} = A\mathbf{y}_{t,k_y} + B\mathbf{u}_{t,k_u} + E\mathbf{v}_{t,k_v}
\end{equation}
where $y_{t+1}\in\mathbb{R}^n$ is the predicted indoor temperature vector at time instant $t+1$, $\mathbf{y}_{t,k_y}\in\mathbb{R}^{nk_y}$ is the stacked historical indoor temperature measurements during time period $[t-k_y+1,t]$, which is defined as $\mathbf{y}_{t,k_y}:=[y_t^{\mathrm{T}},y_{t-1}^{\mathrm{T}},\cdots,y_{t-k_y+1}^{\mathrm{T}}]^{\mathrm{T}}$; similarly, $\mathbf{u}_{t,k_u}\in\mathbb{R}^{mk_u}$ and $\mathbf{v}_{t,k_v}\in\mathbb{R}^{pk_v}$ are the stacked HP power input and ambient climate conditions during time intervals $[t-k_u+1,t]$ and $[t-k_v+1,t]$, respectively; $A,B$ and $E$ are system matrices with appropriate dimensions. The above linear model encompasses a wide range of the prediction models for indoor thermal dynamics developed by black-box approaches, e.g., Auto Regressive with eXogeneous inputs (ARX), and gray-box approaches, e.g., RC-network \cite{wang2019data,drgovna2020all,bacher2011identifying}. 
\subsection{Heat Pump and Ambient Noises}

For HP noise concerns, our primary focus is on the noise generated by the fan in the outside unit. In residential settings, HP fan noise generally ranges from 40-60 decibels, and is typically the main source of HP noise disturbance for nearby residents. As shown in \cite{langerova2025air,sti24}, the noise generated by HPs is a nonlinear and nonconvex function of the HP power, and roughly follows a logarithm-like or sigmoid-like function. However, due to the lack of definitive studies showing that all ASHPs adhere to such a noise pattern, we impose no explicit assumption about the relationship between the HP noise level and power consumption.


In our control scheme design, we only assume the existence of a general noise pattern as defined in \eqref{eq:hp_noise}, which can be derived through theoretical analysis or experimental data:
\begin{equation}\label{eq:hp_noise}
    L^{\text{hp}} = f(P)
\end{equation}
where $L^{\text{hp}}$ is the HP noise level in decibels, $P$ is the HP power input, and $f(\cdot)$ is a function representing the HP noise pattern. The implicit assumption behind the above analysis is that the noise generated by HPs varies with their electrical power consumption, i.e., thermal output, and follows a certain predictable pattern. This assumption requires that the considered HPs should be inverter HPs, whose fan speed can be modulated and power input is adjustable, since the fan speed and power input for ON/OFF controlled HPs are generally fixed .

When defining the acoustic nuisance caused by a HP, another factor that should be considered is ambient noise (background noise), which might be caused by traffic noise, alarms, extraneous speech, animal noise, and more. In this work, we impose no specific pattern for ambient noise, assuming only that the predicted ambient noise levels are accessible. This is a practical assumption, and there are many works available that focus on
developing ambient noise prediction algorithms, see \cite{zhang2020time,renaud2023deep} and references therein.
Thus, without loss of generality, in our upcoming MPC design, we assume that the predicted ambient noise level $L^{\text{amb}}$ within the MPC prediction horizon is available.

\subsection{Control Design Objective}
The main control objective is to mitigate the acoustic nuisance caused by HP operation in the surrounding environment. It should be noted that reducing the acoustic nuisance of HPs does not equate to minimizing absolute HP noise, which would typically mean shutting down HPs. Instead, it involves reducing the relative impact of HP noise compared to ambient noise levels. Through appropriate HP control, the combined noise from the HP and its surroundings should be dominated by the ambient noise, effectively concealing HP noise within it and thereby mitigating acoustic pollution. In addition to reducing noise pollution, the control objective should also consider indoor comfort and energy costs. 

For simplicity in MPC design, our approach does not account for the spectral characteristics of noise signals. Future studies incorporating human factors determining the human-perceived noise nuisance associated with different noise frequencies are warranted.

\section{Model Predictive Control Design}\label{sec:mpc_design}

\subsection{MPC Formulation}
In this subsection, a general MPC problem is formulated to adaptively reduce the effect of HP noise on the environment. An MPC problem achieving our design objective can be formulated as follows
\begin{subequations}\label{eq:general_mpc}
    \begin{align}
        \min_{u_t}&\ \underbrace{\sum_{t=0}^Nl(u_t,y_t)}_{J_o} + \eta \underbrace{\sum_{t=0}^Nh(L_t^{\text{hp}}, L_t^{\text{amb}})}_{J_n}\label{eq:obj}\\
        \text{s.t. }& y_{t+1} = A\mathbf{y}_{t,k_y} + B\mathbf{u}_{t,k_u} + E\mathbf{v}_{t,k_v}, \label{eq:therml} \\
        & L^{\text{hp}}_t = f(u_t),\label{eq:noise}\\
        & y_t\in\mathcal{Y} \text{ and } u_t\in\mathcal{U},\  \forall t \in \{0,\cdots, N\}\label{eq:comfort}
    \end{align}
\end{subequations}
where $N$ is the length of prediction horizon, $J_o$ is total operational cost with $l(u_t,y_t)$ as the stage cost at sampling instant $t$, $J_n$ is the total noise cost within the prediction horizon with $h(L^{hp}_t, L^{\text{amb}}_t)$ as the stage noise cost defined based on HP noise $L_t^{\text{hp}}$ and ambient noise $L_t^{\text{amb}}$, $\eta\geq 0$ is a user-defined weighting factor, constraint \eqref{eq:therml} is the building thermal dynamics defined in \eqref{eq:thermal_dyn}, constraint \eqref{eq:noise} defines the noise pattern of the HP in \eqref{eq:hp_noise}, $\mathcal{Y}$ and $\mathcal{U}$ in \eqref{eq:comfort} are admissible regions of indoor temperature and heat pump power input. This optimization problem defines a general control task to minimize the weighted sum of HP operational cost and HP noise cost while ensuring indoor comfort constraints and HP input constraints.

\textit{Remark 1}: It should be highlighted that the design objective is to mitigate the HP noise nuisance w.r.t. the environment noise. Consequently, the definition of the noise cost function $J_n$ in \eqref{eq:obj} should reflect the relative noise nuisance of the HP, rather than its absolute value. For example, intuitively, when the ambient environment is noisy, the HP can operate with louder noise, possible to achieve higher energy efficiency or lower energy bills, without incurring a high acoustic nuisance. Similarly, when the ambient environment is quiet, even a moderate noise level of HP can lead to more nuisance because the HP noise plays a dominating role in the total noise.

\subsection{Piecewise Affine Approximation of HP Noise Pattern}
Assuming a general HP noise pattern, this subsection presents a piecewise linear approximation of the HP noise pattern and develops a computationally tractable formulation. 

As explained in Section \ref{sec:problem_setting}.B, since the specific noise pattern might vary depending on the individual HP system, we make no explicit assumption about the HP noise pattern, and aim at developing methods that are adaptable to a broader range of HP noise patterns for enhancing the applicability of the proposed approach.

\begin{figure}[htb]
    \centering
    \includegraphics[width=0.8\linewidth]{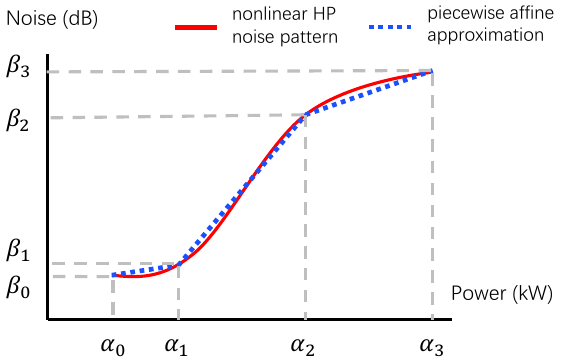}
    \caption{Nonlinear heat pump noise pattern and its piecewise affine approximation.}
    \label{fig:noise_pattern}
\end{figure}

In this work, piecewise affine functions are utilized to approximate the HP noise pattern, which might be nonlinear and nonconvex. Fig. \ref{fig:noise_pattern} shows an example of using three pieces of affine functions to approximate a sigmoid-like noise pattern. To provide a general approximation scheme, we assume that affine functions comprising $k$ pieces are used in HP noise approximation. The admissible scope of HP control input is partitioned into $k$ intervals that are defined by $\alpha = [\alpha_0,\cdots,\alpha_{k}]^{\mathrm{T}}$ with $[\alpha_i,\alpha_{i+1}]$ $(i=0,\cdots,k-1)$ representing one interval, where the HP noise pattern is approximated via a piecewise affine function. Correspondingly, the vector $\beta := [\beta_0,\cdots,\beta_{k}]$ is defined with $\beta_i$ as the HP noise level when its power input $u = \alpha_i$. Then, for any HP power input $u\in[\alpha_i,\alpha_{i+1}]$, there exist real-valued parameters $\lambda_i$ and $\lambda_{i+1}$ such that
\begin{equation}
    u = \lambda_i \alpha_i + \lambda_{i+1}\alpha_{i+1}
\end{equation}
with $0 \leq \lambda_i \leq 1$ and $\lambda_i + \lambda_{i+1} = 1$. Correspondingly, the value of the approximated HP noise level $\hat{L}^{\text{hp}}$ is 
\begin{equation}
    \hat{L}^{\text{hp}} = \lambda_i\beta_i + \lambda_{i+1}\beta_{i+1} 
\end{equation}
To denote the approximated noise pattern within the whole admissible input range  $[\alpha_0,\alpha_{k}]$, binary variables $z_i\in\{0,1\}$ $(i = 1,\cdots,k)$ are introduced with $z_i = 1$ indicating $u\in[\alpha_{i-1},\alpha_{i}]$. Finally, the piecewise affine approximated HP noise pattern can be expressed as the following mixed-integer linear constraints
\begin{subequations}\label{eq:milp_piecewise}
    \begin{align}
        &u = \sum\nolimits_{i=0}^{k} \lambda_i\alpha_i,\\
        &\hat{L}^{\text{hp}} = \sum\nolimits_{i=0}^{k}\lambda_i\beta_i,\\
        & \lambda_{i-1} + \lambda_{i} \leq z_i,\ \forall i \in \{1,\cdots, k\}\\
        & \lambda_i \geq  0, \forall i \in \{1,\cdots, k\} \\
        &\sum\nolimits_{i = 1}^k z_i = 1,\ z_i \in\{0,1\}.
    \end{align}
\end{subequations}
The above mixed-integer linear constraints can replace the HP noise pattern constraint in \eqref{eq:noise}, which might be nonlinear and nonconvex, and enable a universal approximation for various HP noise patterns. The approximation accuracy can be adjusted by $k$ -- the number of affine functions utilized in the approximation. 
\subsection{Noise Cost Function Design}
This subsection presents several possible options for defining the noise cost function $J_n$ in \eqref{eq:general_mpc}. Recall that our control objective is to mitigate the relative acoustic nuisance in comparison to ambient noise, rather than minimizing the absolute HP noise level, so that the HP noise is hidden in the ambient noise. Accordingly, the design of the noise cost $J_n$ should emphasize relative noise mitigation. 

\subsubsection{Option 1}:\label{sec:opt1}
The first option of the noise cost is defined as
\begin{equation}\label{eq:inverse_penalty}
    J_n:=\sum\nolimits_{t=0}^N {L^{\text{hp}}_t}/{L^{\text{amb}}_t}
\end{equation}
The above cost function penalizes HP noise according to the ambient noise level. Higher ambient noise imposes less penalty on HP noise. Consequently, this cost function incentivizes HP to work at a higher load when the ambient environment is noisy and at a lower load in quieter settings.
\subsubsection{Option 2} \label{sec:opt2}
While the cost function defined in \eqref{eq:inverse_penalty} is straightforward and easy to implement, it fails to impose direct regulation on the mixed noise and may not prevent HP noise from dominating the ambient noise. According to the acoustic properties of combined sounds, the sound level of the mixed noise from HP and ambient sources can be calculated as
\begin{equation}\label{eq:mixed_noise}
    L^{\text{mix}} = 10\cdot\log_{10}\left(10^{\frac{L^{\text{amb}}}{10}} + 10^{\frac{L^{\text{hp}}}{10}}\right)
\end{equation}
The above nonlinear function can be used to impose direct constraints on the mixed noise. However, it introduces nonlinear constraints, which might be computationally challenging for certain numerical solvers. 

The definition of the mixed noise level in \eqref{eq:mixed_noise} suggests that when the ambient noise level exceeds the HP noise, the mixed noise level will be primarily dominated by the ambient noise, due to the power function applied to each noise level. Thus, an alternative approach is to penalize instances when the HP noise exceeds the ambient noise, resulting in the following noise cost function definition
\begin{equation}\label{eq:res_penalty}
    J_n := \sum\nolimits_{t=0}^N \delta_t,\quad L_t^{\text{hp}} \leq L_t^{\text{amb}} + \delta_t
\end{equation}
with $\delta_t \geq 0$. The above cost function is equivalent to the nonlinear noise cost function $J_n := \sum_{t=0}^N(L_{t}^{\text{hp}} - L_t^{\text{amb}})^+$, where $x^+$ is defined as $x^+ = x$ if $x\geq 0$ and otherwise $x^+ = 0$. This cost function incentivizes that HP noise does not exceed ambient noise, allowing the mixed noise to be predominantly influenced by ambient sounds, thereby masking HP noise within the background noise.




\textit{Remark 2}: It is worth noting that the mixed-integer linear formulations in \eqref{eq:milp_piecewise} and \eqref{eq:res_penalty} are designed to enhance the applicability and computational feasibility of our approaches for a variety of HP noise patterns and for compatibility with most numerical solvers. However, if the available solvers are capable of handling the specific, possibly nonlinear and nonconvex, HP noise pattern in \eqref{eq:hp_noise} and the mixed noise pressure level definition in \eqref{eq:mixed_noise}, this nonlinear relationship could be directly incorporated into \eqref{eq:general_mpc}, potentially improving control performance.
\begin{figure}[htb]
    \centering
    \includegraphics[width=0.85\linewidth]{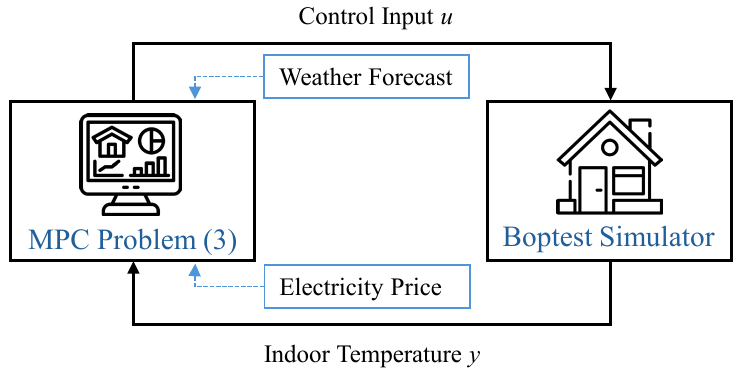}
    \caption{Simulation diagram.}
    \label{fig:simu_diag}
\end{figure}

\begin{figure}[htp]
    \centering
    \includegraphics[width=0.8\linewidth]{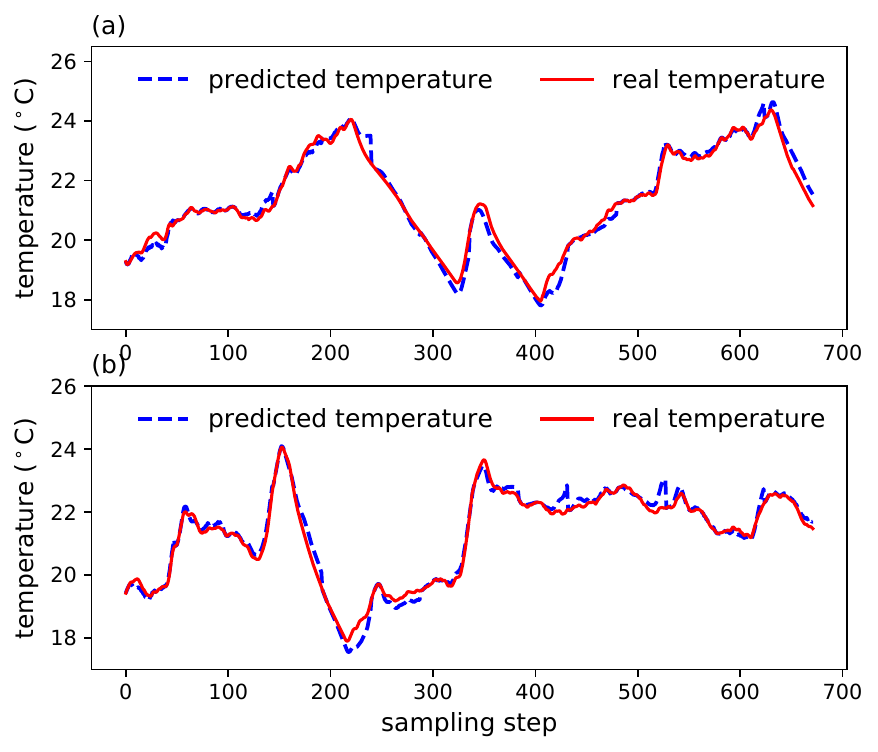}
    \caption{Open-loop prediction performance of ARX model: (a) training set (MAE = 0.16$^\circ$C), (b) test set (MAE = 0.14$^\circ$C).}
    \label{fig:arx_model}
\end{figure}
\section{Simmulation Results}\label{sec:simu_results}
This section presents numerical simulation results to demonstrate the viability and effectiveness of our proposed design framework. 
The building model {\tt bestest\_hydronic\_heat\_pump} in the building control test platform {\tt boptest}~\cite{blum2021building} is utilized as a high-fidelity simulator to test our design. The considered building model is a residential building with a rectangular floor plan $12$m$\times$$16$m, a height of $2.7$m, and an air-to-water HP of 15 kW nominal heating capacity for floor heating. See \cite{blum2021building} for more details about this building control test platform. The diagram of our simulation is shown in Fig. \ref{fig:simu_diag}. At each sampling instant, the HP control input signal $u_t$ is computed by solving \eqref{eq:general_mpc}. With the computed HP control input, the building simulator updates its internal states and return the updated indoor temperature $y_{t+1}$. All simulations are performed on an Intel Xeon W-2223 CPU at 3.60GHz with 16G RAM. MPC problems are modeled via the Python package {\tt gurobipy} and solved using {\tt Gurobi 11.0} \cite{gurobi}. 

\begin{figure}[htb]
    \centering
    \includegraphics[width=0.7\linewidth]{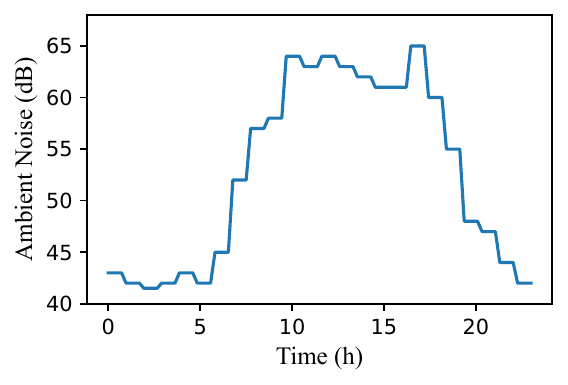}
    \caption{Ambient noise profile used in simulation.}
    \label{fig:noise}
\end{figure}

\begin{figure*}[h]
        \centering
        \includegraphics[width=0.95\linewidth]{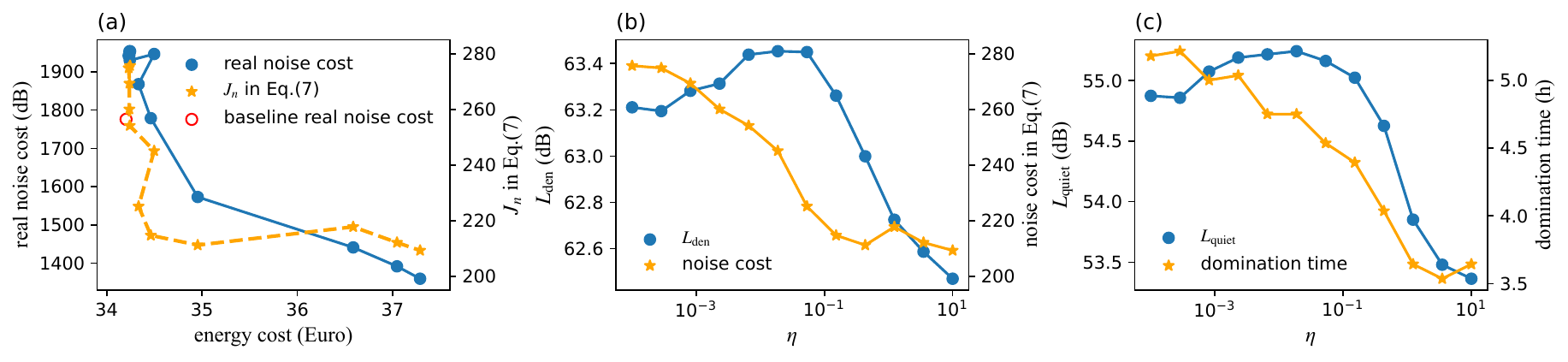}
        \caption{\small Simulation results for noise cost in \eqref{eq:inverse_penalty}: (a) Pareto curves of energy cost and noise cost, (b) $L_{\text{den}}$, (c) $L_{\text{quiet}}$ and domination time.}
        \label{fig:inv_penalty}
\end{figure*}

\begin{figure*}[h]
        \centering
        \includegraphics[width=0.95\linewidth]{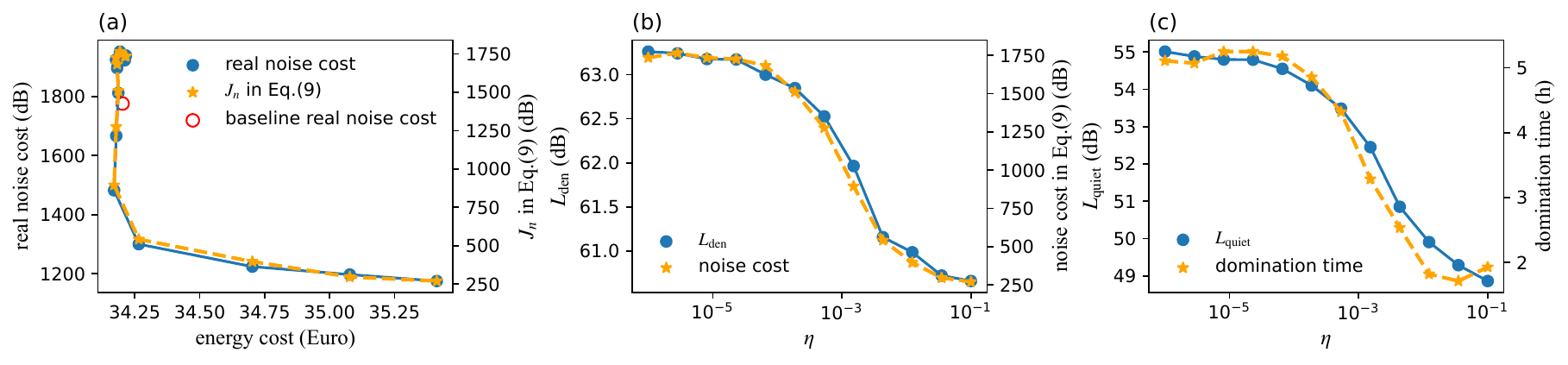}
        \caption{\small Simulation results for noise cost in \eqref{eq:res_penalty}: (a) Pareto curves of energy cost and noise cost, (b) $L_{\text{den}}$, (c) $L_{\text{quiet}}$ and domination time.} 
        \label{fig:res_penalty}
\end{figure*}


{\fullversion In our simulation, the sampling period is selected as $15$ minutes for both prediction model development and MPC design. An ARX model is identified using randomly generated open-loop control signals to approximate the thermal dynamics of the building. Notably, the control signals are forced to be off (on) when the building is overheated (overcooled). The {\tt boptest} simulator utilized in our simulation also involves the internal thermal gains from 5 occupants, but they are not forecast and therefore neglected in our ARX model. The ARX model is with the following structure
\begin{subequations}
    \begin{align*}
        y_t = &\sum_{k=1}^{n_a}a_ky_{t-k} + \sum_{k=1}^{n_b}b_ku_{t-k}+\sum_{k=1}^{n_c} c_kT_{t-k}\\
        & +\sum_{k=1}^{n_d}d_kS_{t-k}
    \end{align*}
\end{subequations} 
where $y_t$ is the indoor temperature at time instant $t$, $T_t$ denotes the ambient temperature, $S_t$ denotes the solar irradiation. The prediction horizon is set as $12$ hours. In our model the parameters $(n_a, n_b,n_c,n_d)$ are set as $(4,1,2,2)$. The parameters $(a_k,b_k,c_k,d_k)$ are identified by minimizing the sum of the squares of the indoor temperature prediction error via {\tt {scipy.optimize.least\_squares}} \cite{scipy}. For training and testing the ARX model, two independent 7-day of datasets are utilized, i.e. 7*24*4 = 672 data points, in both training and test datasets.

Fig. \ref{fig:arx_model} depicts the real indoor temperature profiles and their open-loop predicted values using the ARX model for both the training and test datasets. The mean absolute errors (MAEs) for both training and test sets are 0.16 $^\circ$C and 0.14$^\circ$C, respectively, which implies that the ARX model provides satisfactory prediction performance for subsequent MPC design.}

For design simplicity, the admissible range of HP power input is scaled to such that $u_t\in[0,1]$. The indoor comfort constraint is set as $19^\circ\text{C} \leq y_t \leq 24^\circ\text{C}$. For the HP noise pattern, its real value is assumed to be identical with the piece-wise affine approximated value, that are defined with the vectors $\alpha$ and $\beta$ used in \eqref{eq:milp_piecewise} as
$\alpha = [0,0.2,0.7,1]$ and $\beta = [0,40,60,60]$.
The operational cost function, also referred as energy cost, in the MPC design \eqref{eq:general_mpc} is defined as the electricity cost within the prediction horizon, i.e., $J_o:=P_{\text{max}}\cdot\sum e_tu_t$, where $P_{\text{max}}$ is the maximal HP power input, $e_t$ is the day-ahead electricity price. The prediction horizon $N=32$, i.e., $8$ hours. The ambient noise pattern used in our simulation is shown in Fig. \ref{fig:noise}, which is generated based on the results in \cite{thomas2018evaluation,zhang2020time}. It can be seen that the environment is quiet during the early morning, evening and night hours, and is noisy at noon and in the afternoon, which is consistent with our everyday experience.

In our case studies, the proposed two options of noise cost $J_n$ defined in \eqref{eq:inverse_penalty} and \eqref{eq:res_penalty} are tested for the MPC design, respectively. For the sake of brevity, the explicit MILP formulations for \eqref{eq:general_mpc} with HP noise pattern approximation \eqref{eq:milp_piecewise}, and the noise cost functions \eqref{eq:inverse_penalty} and \eqref{eq:res_penalty} are formulated as follows:
{\fullversion \begin{subequations}\label{eq:total_opt}
    \begin{align}
        \min_{\substack{u_t,\lambda_{i,t},z_{i,t}\\\delta_t, \hat{L}^{hp}_t}}\ & P_{\text{max}}\cdot\sum\nolimits_{t=0}^Ne_tu_t + \eta\cdot\sum\nolimits_{t = 0}^N\delta_t \\
        \text{s.t. }& y_{t+1} = A\mathbf{y}_{t,k_y} + B\mathbf{u}_{t,k_u}+E\mathbf{v}_{t,k_v},\\
        & u_t = \sum\nolimits_{i=0}^{k}\lambda_{i,t}\alpha_i,\ \hat{L}^{\text{hp}}_t = \sum\nolimits_{i=0}^k\lambda_{i,t}\beta_i,\\
        &\lambda_{i-1,t} + \lambda_{i,t} \leq z_{i,t},\quad\forall i \in \{1,\cdots, k\},\\
        &\lambda_{i,t} \geq 0,\quad \forall i \in \{0,\cdots, k\},\\
        &\sum\nolimits_{i=1}^k z_{i,t} = 1, \quad z_{i,t}\in\{0,1\},\\
        & 19 \leq y_t \leq 24,\quad 0\leq u_t \leq 1,\\
        & \hat{L}^{\text{hp}}_t \leq L_{t}^{\text{amb}} + \delta_t , \quad  \delta_t \geq 0,\label{eq:extra}\\
        & \forall t \in \{0,\cdots, N\}.
    \end{align}
\end{subequations}
where the operation cost $J_o$ is defined as $P_{\text{max}}\sum_{t=0}^Ne_tu_t$.
In case of the noise cost function defined in \eqref{eq:inverse_penalty}, the term $\sum_{t=0}^N\delta_t$ in the above cost function is replaced by $\sum_{t=0}^N\hat{L}^{\text{hp}}_{t}/L^{\text{amb}}_t$, and constraint \eqref{eq:extra} is removed.}

In the simulation, different values of $\eta$ in \eqref{eq:obj} are tested. For each value of $\eta$, seven days of closed-loop simulation are performed. Based on the simulation results, the total noise cost that are defined in \eqref{eq:inverse_penalty} and \eqref{eq:res_penalty}, and energy cost (electricity consumption defined in \eqref{eq:total_opt}) during the simulation period are computed. Besides, since the noise costs defined in \eqref{eq:inverse_penalty} and \eqref{eq:res_penalty} are the convex approximations of the desired noise penalty, which is to ensure that the mixed noise is primarily dominated by ambient noise, we introduce a common metric: 
\begin{equation}\label{eq:real_cost}
\textit{real noise cost }:=\sum\nolimits_t L^{\text{mix}}_t - L^{\text{amb}}_t
\end{equation}
to evaluate different approaches. In addition, the daily-averaged values of the following metrics are also evaluated:
\begin{itemize}
    \item $L_{\text{den}}$ of mixed noise: the day-evening-night noise level. $L_{\text{den}}$ is used to measure the overall sound exposure over 24 hours. It is defined as the equivalent sound level with different penalties over different time periods in day, evening and night \cite{lden}. 
    \item $L_{\text{quiet}}$ of mixed noise: $L_{\text{quiet}}$ is defined as the equivalent sound level during quiet time (10:00 pm - 7:00 am).
    \item \textit{domination time}: the total time over 24 hours that the mixed-noise level is dominated by the HP noise. 
\end{itemize}
Furthermore, the \textit{baseline} approach, in which the HP is operated to minimize the energy cost while complying with Switzerland's day-night noise regulations ($60$ dB limit during daytime and $50$ dB limit at night \cite{ehpa20}), is also considered in our case study.

\begin{table}[htb]
    \centering
    \caption{Performance summary with different noise cost functions.}\resizebox{\linewidth}{!}{
    \begin{tabular}{l|c|c}\toprule\hline
    & noise cost in \eqref{eq:inverse_penalty} & noise cost in \eqref{eq:res_penalty} \\\hline
        noise cost $J_n$ reduction percentage (\%)&  24.09 & 84.48\\
        \textit{real noise cost} reduction percentage (\%) & 30.43&39.38 \\
        energy cost increase percentage (\%)& 8.89 & 3.50 \\
        $L_{\text{den}}$ reduction (dB) & 0.74 & 2.60 \\
        $L_{\text{quiet}}$ reduction (dB) &1.51 & 6.15 \\
        \textit{domination time} reduction (h) &1.54 & 3.39 \\
        average MPC computation time (s) & 2.17 & 0.87  \\
        \hline
    \end{tabular}} 
    \label{tab:performance_summary}
\end{table}

Simulation results are plotted in Fig. \ref{fig:inv_penalty} and \ref{fig:res_penalty}. Table \ref{tab:performance_summary} summarizes the results in terms of the maximal noise cost reduction percentage and the corresponding \textit{real noise cost} reduction percentage, energy cost increase percentage, $L_{\text{den}}$ reduction, $L_{\text{quiet}}$ reduction, and domination time reduction for all considered values of $\eta$ with both noise cost definitions in \eqref{eq:inverse_penalty} and \eqref{eq:res_penalty}, respectively.

Fig. \ref{fig:inv_penalty} presents the simulation results using the noise cost function $J_n$ defined in \eqref{eq:inverse_penalty}. The Pareto curves of the \textit{real noise cost} in \eqref{eq:real_cost} and the noise cost $J_n$ in \eqref{eq:inverse_penalty} w.r.t. energy cost in Fig. \ref{fig:inv_penalty}(a), along with Table \ref{tab:performance_summary}, indicates that noise cost $J_n$ can be reduced by $24.09\%$ with an $8.89\%$ increase in energy cost. In the meanwhile, the \textit{real noise cost} is reduced by $30.47\%$. Fig. \ref{fig:inv_penalty}(b) illustrates the variations in $L_{den}$ and noise cost as $\eta$ in \eqref{eq:obj} increases. It is observed that while both $L_{\text{den}}$ and noise cost $J_n$ generally follow a downward trend, their patterns are not entirely consistent, implying that a reduction in noise cost does not necessarily correlate with decreased noise nuisance in $L_{\text{den}}$, as also discussed in Section \ref{sec:opt2}. Similarly, the inconsistency for $L_{\text{quiet}}$ and domination time is also visible in Fig. \ref{fig:inv_penalty}(c).

In Fig. \ref{fig:res_penalty}(a), the Pareto curves illustrate the trade-off between noise and energy costs for the noise cost function in \eqref{eq:res_penalty}. Together with Table \ref{tab:performance_summary}, it can be observed that the noise cost $J_n$ and the \textit{real noise cost} are reduced by $84.48\%$ and $39.38\%$, respectively, with only a $3.50\%$ increase in energy cost. Fig. \ref{fig:res_penalty}(b) presents the values of $L_{\text{den}}$ and noise cost $J_n$ across various values of $\eta$. Similarly, as depicted in \ref{fig:res_penalty}(c), both $L_{\text{quiet}}$ and \textit{domination time} decreases as $\eta$ increases, achieving a notable 6 dB reduction in $L_{\text{quiet}}$ and 3.39 h reduction in \textit{domination time}, which are much larger than the case with $J_n$ defined in \eqref{eq:inverse_penalty}, where a reduction of $1.51$ dB in $L_{\text{quiet}}$ and $1.54$ h in \textit{domination time} are achieved. Notably, Fig.\ref{fig:res_penalty} shows a much more consistent pattern among the noise cost $J_n$ in \eqref{eq:res_penalty}, \textit{real noise cost}, $L_{\text{den}}$, $L_{\text{quiet}}$ and \textit{domination time} than in Fig. \ref{fig:inv_penalty} where the noise cost \eqref{eq:inverse_penalty} is considered.  This implies that the noise cost in \eqref{eq:res_penalty} is more effective in penalizing the mixed noise level and ensuring the mixed noise is dominated by the ambient noise.

In addition, as shown in Fig. \ref{fig:inv_penalty}(a) and Fig. \ref{fig:res_penalty}(a), the \textit{baseline} approach mainly aims at reducing the energy cost and cannot consider the ambient noise profile, which leads to large noise costs. Besides, one might notice that in Fig. \ref{fig:inv_penalty}(a) and Fig.\ref{fig:res_penalty}(a) the energy cost does not always increase as the noise cost decreases, which is possibly due to modeling errors causing the HP to deviate from the predicted optimal value for maintaining indoor comfort. This issue could be mitigated by using stochastic or robust optimization-based approaches to enhance the robustness of the MPC solution.


\section{Conclusions}\label{sec:conclusion}
This paper presents the first investigation into HP noise mitigation within the context of building climate control. The proposed approach extends the standard economic MPC design for building climate control to incorporate HP noise reduction. By adopting piecewise linear approximation, the proposed approach effectively accommodates diverse HP noise patterns while maintaining computational efficiency by solving MILP problems. The proposed noise cost functions incentivize that the HP noise does not dominate the ambient noise, thereby reducing its acoustic impact on surrounding environments. Simulation results using a high-fidelity building simulator demonstrate that, with the proposed MPC design, HP noise can be mitigated with only a minor increase in energy costs.   

{\fullversion Future works include analyzing the influence of HP noise mitigation on its energy efficiency and experimentally investigating the viability and effectiveness of the proposed approach.}

\bibliographystyle{IEEEtran.bst}

\bibliography{ref} 

\end{document}